\documentclass[final,5p,times,twocolumn]{elsarticle}

\usepackage{graphicx}
\usepackage{amssymb}
\usepackage{mathrsfs}
\usepackage[normalem]{ulem}

\usepackage{lineno}

\usepackage{amsmath,amsfonts,amsthm} 
\usepackage{hyperref}
\usepackage{color}
\usepackage{booktabs}

\newcommand\numberthis{\addtocounter{equation}{1}\tag{\theequation}}

\biboptions{square}

\bibpunct{(}{)}{;}{a}{}{,} 

\journal{Journal Name}

\begin{document}

\begin{frontmatter}


\title{Astroalign: A Python Module for Astronomical Image Registration}



\author[utrgv]{Martin Beroiz}
\author[cifasis,iate]{Juan B. Cabral}
\author[duke]{Bruno Sanchez}

\address[utrgv]{
   University of Texas Rio Grande Valley (UTRGV), Texas, USA}
\address[cifasis]{
   Centro Internacional Franco Argentino de Ciencias de la
   Informaci\'on y de Sistemas (CIFASIS, CONICET--UNR),
   Ocampo y Esmeralda, S2000EZP,
   Rosario, Argentina.}
\address[iate]{
   Instituto de Astronom\'ia Te\'orica y Experimental -
   Observatorio Astron\'omico C\'ordoba (IATE, UNC--CONICET),
   C\'ordoba, Argentina.}
\address[duke]{Department of Physics, Duke University, 120 Science Drive, Durham, NC, 27708, USA}

\begin{abstract}
We present an algorithm implemented in the \emph{Astroalign} Python module for image registration in astronomy.
Our module does not rely on WCS information and instead matches three-point asterisms (triangles) on the images to find the most accurate linear transformation between them.
It is especially useful in the context of aligning images prior to stacking or performing difference image analysis.
Astroalign can match images of different point-spread functions, seeing, and atmospheric conditions.
\end{abstract}

\end{frontmatter}



\section{Introduction}
Image registration is the process of transforming images that use different coordinate systems, so that after the transformation they share a common frame.

Image registration is widely used in the medical fields \citep{Fischer_2008}, including nuclear imaging \citep{Hutton2002} and radiology \citep{Hill_2001}.
It is also widely used in Computer Vision \citep{ZITOVA2003977} for object recognition \citep{objectrecon}, image stitching \citep{stitching} and visual mapping \citep{visualmapping}.
Image registration is used in virtually any field that needs to reconcile images taken from different points of view.
In astronomy, image registration is used for tasks such as stacking images, difference image analysis, and creating mosaics.

In the particular field of astronomy, all images have a natural common reference frame, given by the spherical coordinates of position on the celestial sphere.
Registration between two images usually amounts to finding the mathematical transformation between each coordinate frame.

The astrometric information of an astronomical image is typically stored in the header of a FITS file using a World Coordinate System (WCS) \citep{wcs}. 
Solving for the WCS is only one step away from registering any two images, since the last remaining step
is to remap one or both images into the desired destination reference frame.
The Python packages `reproject' \citep{reproject} and `MontagePy' \citep{Montage_2009, Montage_2017} work in this way.

When no WCS information is available, and the pointing of the image is uncertain or unknown,
a program like astrometry.net \citep{astrometrynet2010} will identify the portion of the sky and generate appropriate WCS information.

Generating WCS from a completely unknown image is a daunting challenge, one that astrometry.net handles handsomely well.
But despite its undisputed usefulness, it may be an overkill for some cases.
Roughly speaking, the task is to spot a particular section of the sky on 40K square degrees for images that typically span less than 1 square degree.
This exhaustive search can be computationally expensive and it requires the download of several hundreds of megabytes of index files\footnote{astrometry.net has an online tool that does not require downloading files.}.

Unlike Astrometry.net, Astroalign\footnote{\href{http://github.com/toros-astro/astroalign}{http://github.com/toros-astro/astroalign}} does not try to solve for WCS coordinates.
Astroalign is a fast and lightweight Python module that registers astronomical images bringing the images to a common frame via the best linear transformation between them.
For this reason it is particularly suited for situations where we want to correct for drift or displacements over an observation night, or if we want to reconcile two images taken by different observatories, but where no astrometry is needed.

Astroalign assumes little about the origin and format of the image, making it suitable for astrophotographers and amateur astronomers, who do not usually deal with FITS-formatted images.
Regardless, Astroalign is compatible with modern astronomy data objects such as astropy's NDData, ccdproc's CCDData, and Numpy arrays (with or without masks), so that astronomers can easily integrate with other Python packages.
Astroalign can match images of different point-spread functions, seeing and atmospheric conditions.
Astroalign is tested under continuous integration and has documentation available online\footnote{\href{https://astroalign.readthedocs.io}{https://astroalign.readthedocs.io}}.
It may not work, or work with special care, on images of extended objects with few point-like sources or in crowded fields.



\section{The Algorithm}
\label{algo}

In the field of Computer Vision (CV), generic feature-based registration routines like SIFT \citep{sift}, SURF \citep{surf}, or ORB \citep{orb}, try to identify and match ``interesting points'' using corner detection routines to estimate a point correspondence between the images.
These generally fail for stellar astronomical images, since stars are effectively point sources that have very little distinct structure and hence, indistinguishable from each other.
Asterism matching is more robust and closer to the way humans match images.

The idea of using star patterns to match stellar images goes back to an early publication by \cite{trianglematching}. The author deals with the very similar problem of finding the correspondence between two lists of coordinates from star catalogs. His method already contains many of the core ideas used in our method, but it is prone to generate a large number of outliers, difficult to remove. It also scales poorly with the fourth power of the number of reference points.

\cite{PalBakos2006} and \cite{Tabur2007} extend the ideas in \cite{trianglematching} to large catalogs of sources and include possible non-linear distortions of the field. 
Astrometry.net \citep{astrometrynet2010} also improves on many of the shortcomings from \cite{trianglematching}. Astrometry.net uses quadrilaterals instead of triangles and a Bayesian decision criteria to pick the correct image transformation.
We discuss comparison with previous methods in Section \ref{sec:comp}.


Similar to previous work, the core idea of Astroalign's algorithm consists of characterizing asterisms by using ``geometric hashes'' that are invariant to translation, rotation, and scaling. Similar asterisms will have similar invariant tuples in both images so a correspondence between them can be made.

As an example, the lengths of the sides of a triangle are invariant to translation and rotation. They remain the same whatever position or orientation the triangle might have. Additionally, any function of the {\em ratio} of the sides will be invariant to scaling. We explain the invariant tuple that we use in Section \ref{sec:invfeat}.

The idea for the Astroalign algorithm can be summarized in a few steps that we enumerate below.

\begin{enumerate}
\item Do the following for both images \begin{description}
\item Make a catalog of a few brightest sources and push them to a 2D tree (a \emph{k-d tree} data structure for k=2) to quickly query for close neighbors.
\item For each star, select its four nearest neighbors.
\item Create all the $\binom{5}{3}=10$ possible triangles from that set of stars.
\item For each triangle in that set, calculate a tuple of invariants that fully characterize the triangle and push the invariant tuple into another k-d tree.
\item Remove duplicate invariant tuples that come from considering repeated triangles.
\end{description}
\item Identify and match all possible invariants from both lists by querying the nearest neighbor in the k-d trees, thus creating a correspondence between triangles on the images.
\item For each triangle match, make a correspondence between the vertices by considering the lengths of the sides to which they belong.
This way, we establish a point-to-point correspondence for three points.
\item Propose a transformation derived from one point-to-point triangle correspondence. If the proposed transformation fits over 80\% (or 10, whichever is smaller) of the rest of the points, accept the transformation and return the transformed image.
\item If the transformation cannot fit other triangles, or falls below the acceptance threshold, proceed with the next invariant match until one acceptable transformation is found.
\item If the number of unsuccessful tries exceeds a maximum limit, an exception is raised.
\end{enumerate}

In the following sections, we explain in more detail the steps above. 

\subsection{The Image Transformation}
\label{sec:transf}

At the crux of the registration problem is the determination of the geometric pixel transformation that carries one image into another.
In Computer Vision these transformations are called \emph{homographies}.
A homography is a linear transformation between images on projective spaces.
The most general homography can relate images of a planar surface with different perspective points of view.
In our particular case where, for all practical purposes, the sources are located at infinity, we do not use perspective deformations parameters,
but we still adopt the language of CV, due to the widespread use of this formalism in the available software packages. 
The most notable difference with regular 2D matrix operators is the extension of $(x, y)$ coordinates to \emph{homogenous} coordinates $(x, y, 1)$ to include rotations and translations in the same operation.

The parameters for the most general transformation that we consider consists of
a clockwise rotation angle $\alpha$,
a uniform scaling parameter $\lambda$,
and a 2D translation vector $(t_x, t_y)$.
They are summarized in the following coordinate transformation matrix:

\begin{align}
\label{eq:transf}
\left(
 \begin{array}{lll} 
 \lambda \cos \alpha & \lambda \sin \alpha& \lambda t_x \\
 - \lambda \sin \alpha & \lambda \cos \alpha& \lambda t_y \\
 0 & 0 & 1
 \end{array}
\right)
\equiv
\left(
 \begin{array}{lll} 
 a_0 & b_0 & c_0 \\
-b_0 & a_0 & c_1 \\
 0 & 0 & 1
 \end{array}
\right)
\end{align}

To linearize our problem, instead of considering $\{\lambda, \alpha, t_x, t_y\}$ as free parameters, we consider $\{a_0, b_0, c_0, c_1\}$ instead, as if they were independent.

Each correspondence of points $(x_1, y_1) \rightarrow (x_2, y_2)$ in both images is related by equation (\ref{eq:system}),

\begin{align} 
\label{eq:system}
\left(
 \begin{array}{lll} 
 a_0 & b_0 & c_0 \\
-b_0 & a_0 & c_1 \\
 0 & 0 & 1
 \end{array}
\right)
\left(
 \begin{array}{l} 
 x_1 \\
 y_1 \\
 1
 \end{array}
\right)
=
\left(
 \begin{array}{l} 
 x_2 \\
 y_2 \\
 1
 \end{array}
\right).
\end{align}

With three of these point-correspondences, we can solve for the parameters $\{a_0, b_0, c_0, c_1\}$ and $\{\lambda, \alpha, t_x, t_y\}$.


\subsection{Asterisms and Invariant Features}\label{sec:invfeat}

The search for the point correspondences is not done on the set of points itself, but rather uses triangles, or point triplets as proxies.
Unlike single points, triangles can be uniquely characterized.

\begin{figure}
   \centering
   \includegraphics[width = \linewidth]{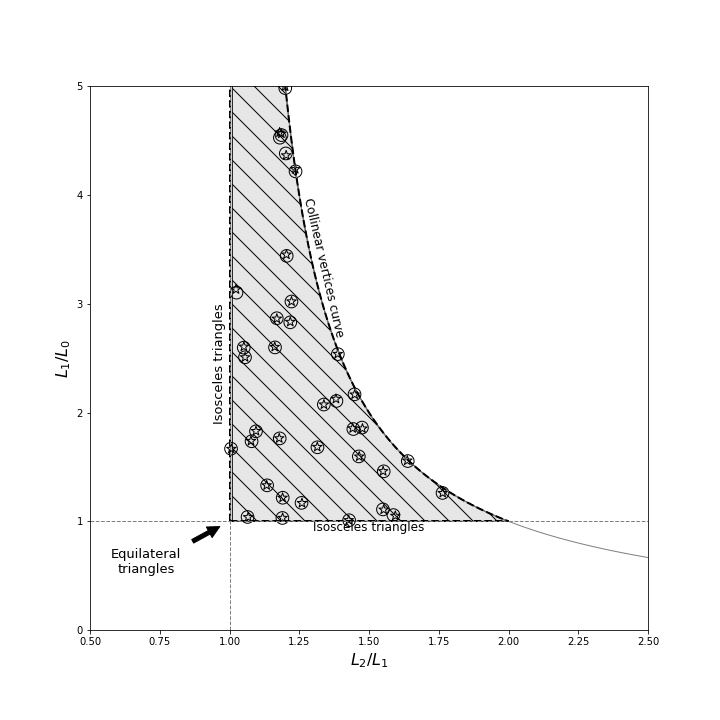}
   \caption{{\bf Region for the Astroalign invariant mapping}. In the region, we graph the invariant tuples from the example in Section \ref{example}. Star-shaped markers represent invariant tuples from the source image and circles represent invariant tuples from the target image.}
   \label{fig:inv_region}
\end{figure}

For a triangle, full knowledge of all three side lengths $\{L_i\}$ is enough to fully characterize it, irrespective of orientation or position.
If we want to characterize it up to a global scaling, then knowing 2 inner angles is sufficient.
Equivalently, knowing two independent ratios of the side lengths is also sufficient.
In fact, any function of two independent length ratios will suffice to fully characterize the triangle.

For Astroalign we chose the mapping defined by: 

\begin{align}
\mathscr{M}(\{L_i\}) &= \left( \frac{L_2}{L_1}, \frac{L_1}{L_0} \right)  \numberthis \label{eq:invdef} \\ 
  \text{where} \; & L_2 \ge L_1 \ge L_0. \label{eq:invdef2}
\end{align}

$\mathscr{M}$ maps into an invariant tuple that fully describes a triangle up to translation, rotation, scaling, and coordinate flipping.
The invariant map $\mathscr{M}$ maps the positive octant of $\mathbb{R}^3$ of all possible side lengths of a triangle
onto a region of the positive quadrant of $\mathbb{R}^2$ in the invariant-feature space.

To find out what this region is, we note that, by definition, $L_2 \ge L_1 \ge L_0$,

\begin{align*}
x &=  \frac{L_2}{L_1} \ge 1 \\
y &=  \frac{L_1}{L_0} \ge 1
\end{align*}

Furthermore, from the triangle inequality:

\begin{align*}
L_2 \leq L_1 + L_0 \implies & x \leq 1 + \frac{1}{y} \\
& y \leq \frac{1}{x-1}
\end{align*}

The curve $y = (x-1)^{-1}$ corresponds to collinear vertices.
Equilateral triangles will map to the point $(1,1)$ and an isosceles triangle will map either to the $x=1$ or $y=1$ line depending on whether the unequal side is the largest or smallest side.
Very peaky triangles will tend to accumulate between the collinear vertices curve and the $x=1$ line for large values of $y$.
These observations can be summarized in Figure \ref{fig:inv_region}, where we also show some example invariant tuples from Section \ref{example}.



\subsection{Selection of Sources}

The selection of stars from Step 1 is an important part of the algorithm.
To identify the positions and fluxes of sources, Astroalign relies on the external package \textit{Source Extraction and Photometry (SEP)} \citep{barbary2016sep},
a project that extracts the core functionality of the program SExtractor \citep{bertin1996sextractor} into a C library and adds a Python API on top.

Mimicking the way humans match images, we pick for the transform estimation, those point sources that have higher chances to be persistent across the images (i.e. the brightest).
Selecting too few sources would make the matching algorithm very sensitive to outliers and missing stars and, thus, unstable. 
Selecting too many sources will make computations unnecessarily complicated and prone to fail because of spurious mismatches.
We found that capping the number of sources at the brightest 50 on both images is a good empirical compromise between robustness and efficiency.

However, the cap at 50 sources does not restrict its use for fields with fewer sources in it. Astroalign is designed to work with as few as three sources---the minimum required to derive a transformation.

In the cases where the source selection requires special care, like in the case of faint sources or very crowded fields,
the source selection can be done separately by the user and the resulting catalogs fed to Astroalign in place of the array images.

\subsection{Triangulation}

Once the catalog of reference sources is made on each image, we push all entries to a k-d tree.
A k-d tree is a data structure optimized to efficiently query for nearest neighbors in $\mathcal{O}(\log n)$ operations\footnote{On average.}.

As mentioned before, we use triangles as a proxy for point-to-point correspondence (see Section \ref{sec:invfeat}).
Given $N$ stars, there are $N(N-1)(N-2)/6$ possible triangles in a fully-connected triangulation.
However, not all of these are needed in practice.

Instead of working with a fully-connected triangulation we adopt a \emph{nearest neighbor triangulation}. With each star and its four nearest neighbors, we calculate the invariant tuples for the 10 triangles that can be made from those five sources. Each invariant tuple is then pushed to another k-d tree.

At the end of this step we have two k-d trees from each image.
The triangles that correspond to the same triplet of stars in the two images, will lie very close on the invariant space.
Consequently, we can query the triangles on one of the k-d trees for its closest partner on the other image.
This is also done efficiently in $\mathcal{O}(n \log n)$ operations.

Every triangle that has a partner in the other image less than 0.1 units away (about 5\% error) is considered a matched triangle.

One triangle match is enough to completely characterize the four parameters of the transformation (\ref{eq:transf})--(\ref{eq:system}),
and this proposed transformation is then tested using the rest of the triangle correspondences.

\subsection{Testing the Transformation}
\label{sec:testtrans}

To estimate how well a transformation fits, we make use of the formalism commonly used in \emph{stereovision} \citep{Zhang1998} to estimate the error of a reprojection.

The reprojection error (RE) is the geometric distance between a projected point and a measured point.
Given two points in homogenous coordinates $x$ and $x'$, and a transformation $H$, the reprojection error is defined as the the distance $d$ from a point ${x'}_i$ to its corresponding epipolar line ${l'}_i = H {x}_i$.

\begin{equation}
d({x'}_i, {l'}_i) = \frac{{x'}^{T}_{i}{l'}_i}{\sqrt{{l'}_1^2 + {l'}_2^2}}
\end{equation}

For a set of correspondence points $\{(x_i, {x'}_i) \}$ we define a global error (\emph{linear criterion}) as:

\begin{equation}
RE = \sum_{i} \left( {l'}_1^2 + {l'}_2^2 \right) d^2({x'}_i, {l'}_i)
\end{equation}

The full formalism of epipolar geometry is quite extensive (see \citet{Zhang1998} for a detailed review of this topic).
We only use it to have a convenient way to estimate the error for a given transformation and set of correspondences.

The actual implementation is taken from the \textit{scikit-image} Python package, that gives an unbiased and fast approximation to the linear criterion
called the ``Sampson distance'' \citep{Luong1993, Fathy2011}.
For each triangle correspondence we calculate the residuals of the transformation with this distance, using the three correspondence points.
If the Sampson distance is smaller than 3, the correspondence is accepted, and rejected otherwise.
A transformation that matches 80\% of the triangle matches or 10 (whichever is lower) is accepted and returned.

This selection process is carried out within a RANSAC process.
The Random Sample Consensus or RANSAC \citep{ransac}, is an iterative method to estimate fit parameters of a model when the data to fit contains a significant fraction of outliers.
It was first introduced as a solution to the ``Location Determination Problem'' to determine the 3D location from where an image was taken.
RANSAC is also distributed as part of the popular CV library \textit{OpenCV} \citep{opencv}.

As is common practice in CV, Astroalign uses a modified implementation of RANSAC\footnote{Initially developed by Andrew Straw.} to estimate the parameters of the homography, while simultaneously ignoring outliers caused by orphan asterisms.


\subsection{Image Reprojection}

Once a transformation is accepted,
the RANSAC algorithm re-calculates the transformation using all of the in-lier triangle vertices identified in the previous step.

Astroalign then uses this re-calculated transformation
to perform a bi-cubic polynomial interpolation of the source image using the \texttt{warp} function from Scikit-Image's \texttt{transformation} module.
A footprint mask of the transformation is also returned along with the transformed image granting the user control on the un-transformed, out-of-boundary pixels.
%
%
%

We tested flux conservation for this type of reprojection, using a simple aperture photometry calculation before and after the image had been transformed.
We found that the ratio of flux per unit area variability is consistent with a Gaussian distribution centered at one, with a variance of ${\sim} 5\times 10^{-3}$.
This means that rigid transformations can introduce errors at the centi-magnitude level, or less than half a percent in flux in the worst-case scenario.
%
We consider that this quantifiable error
is acceptable for the purpose of Astroalign.

\section{Comparison with Other Methods} \label{sec:comp}

Astroalign follows the main ideas developed in \citet{astrometrynet2010} and incorporates several of the existing concepts found in the field of CV.
Astroalign can also be compared to the ideas introduced in other implementations like \citet{PalBakos2006} and \citet{Tabur2007}.

One main difference between our algorithm and Astrometry.net is that Astroalign is not doing a `blind astrometric calibration' nor returns WCS information.
Unlike \citet{trianglematching}, \citet{PalBakos2006} and \citet{Tabur2007},
our main goal is not to cross-match lists of stars, but to find a geometric transformation between a pair of images.
In fact, by design, we restrict the number of sources in our list of guiding stars to a minimum.
We also only consider rigid (linear) transformations between images
and we do not consider non-linear deformations of wide fields like in \citet{PalBakos2006}.

Another difference with previous methods is our triangulation method.
Astrometry.net deals with catalogs of the full sky and uses quadrilaterals at different scales instead of triangles.
The quadrilaterals are chosen to fit in a subdivision of the total sky into Healpix grid cells.
\citet{Tabur2007} prioritizes its search to very acute-angle triangles that are easier to identify in the invariant space.
\citet{PalBakos2006} uses a Delaunay triangulation but adding more redundancy in close-by subgroups of stars to make the triangulation robust to missing or new sources on the field.
Astroalign follows a similar approach than that of \citet{PalBakos2006} by calculating an interlaced full triangulation in small subgroups instead of storing the full triangulation of the field of view.
Our choice to store the triangles in a k-d tree data structure allows us to retrieve similar matches in a fast $\mathcal{O}(n\log{}n)$ number of operations.

Our system to test a transformation is similar to that of \cite{astrometrynet2010} and the \emph{Optimistic Pattern Matching} described in \citet{Tabur2007},
in that we test each hypothetized transformation immediately in the belief that the transformation will most likely be correct.
If the transformation accurately describes other triangles on the mesh, the transformation is immediately accepted and returned.
But unlike Astrometry.net, which uses a Bayesian decision process to verify the prediction power of the position of other stars in the field, we use the reprojection error described in Section \ref{sec:testtrans}.
In our case, the quick estimation and early exit are actually a consequence of using a RANSAC algorithm implementation to search for the transformation.
RANSAC also deals with the problem of missing stars or transient objects in the field.

This short comparison does not pretend to be an exhaustive review of the literature. For a review on earlier work, please refer to \citet{Murtagh1992}.
These differences are summarized in \ref{appendix:a}.


\section{Technical details about the \textit{Astroalign} package}

\subsection{Public API and application example}
\label{example}

The Astroalign Python package splits the registration functionality into a few main functions. The most important ones are:

\begin{description}
    \item[\texttt{find\_transform($\ldots$)}] Estimate the transform between source and target images and also return a tuple with two ordered lists of star correspondences.
    
    \item[\texttt{apply\_transform($\ldots$)}] Apply a given transformation (normally found with the \texttt{find\_transform} function) to some source image. 
    
    \item[\texttt{register($\ldots$)}] Find and apply the transformation to \emph{source} to make it coincide pixel-to-pixel with \emph{target}. This is a convenience wrapper around the two previous functions. 
    The main difference is that \texttt{apply\_transform()} accepts a transformation to apply, but \texttt{register()} always uses the best one found by \texttt{find\_transform()}. This make this function easier to use, but less flexible than the other two combined.
\end{description}

\begin{figure}
   \centering
   \includegraphics[width = \linewidth]{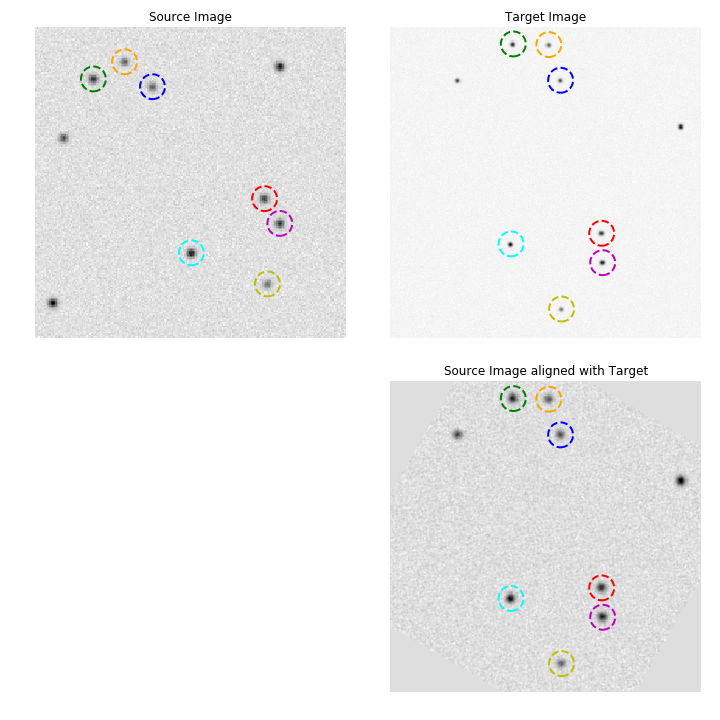}
   \caption{\label{fig:transformations} {\bf Simulated source and target images}.
   On the left panel is the source image with 10 simulated stars.
   The right upper panel is the target image with greater resolution, rotated $30^{\circ}$ clockwise with respect to the source.
   On the bottom panel is the source image aligned to target.
   The star correspondence is marked on a few stars with color-coded circles across all figures and was done using
   \texttt{find\_transform()}.}
\end{figure}

As an application example showing the functionality of the implementation, 
let us consider two sample simulated images as shown in Figure~\ref{fig:transformations}.

The source image on the left panel is $200 \times 200$ pixels, with 10 stars of Gaussian profile
uniformly distributed across the image.
The target image on the right panel simulates an image of the same region of the sky,
but rotated $30^{\circ}$ clockwise with respect to the source image, $300 \times 300$ pixels (dilation factor of 1.5 with respect to source), and containing the same 10 stars of Gaussian profile but with a different full width at half maximum (FWHM).
Notice that the target image has better resolution (and seeing) than the source image.
The parameters of the simulation are summarized in Table \ref{tab:simul}.

\begin{table}
\noindent \centering
\centering
\caption{Image simulation parameters} \label{tab:simul}
\begin{tabular}{lll}
\toprule
                             & Source   & Target   \\ 
\midrule
Number of point sources      & 10       & 10       \\ 
FWHM [pixels]                & 4.71     & 3.53     \\ 
Background level [counts]    & 0.5      & 0.5      \\ 
Background std-dev [counts]  & 0.71     & 0.71     \\
Flux of stars range [counts] & 200--500 & 200--500 \\ \hline
\bottomrule
\end{tabular}
\end{table}

On the third panel of Figure \ref{fig:transformations}, we can see the registered image after \texttt{register()} is applied to the source.
The registered image is now $300 \times 300$ pixels in shape but otherwise aligned with target.

The color circles are some of the stars detected by the function \texttt{find\_transform()},
where the same color identifies the same star on each image.

The invariant tuples internally calculated in this example for both images
are graphed in Figure \ref{fig:inv_region}.
These invariants are not accessible through the public API but are shown to illustrate the method.

One thing to note is that some invariants contain collinear vertices and the distribution over the region is fairly sparse. 
Sparseness is a desired quality that helps in the identification phase.

\subsection{Benchmarks}

To evaluate the performance of Astroalign, we create a collection of random \textit{source} and \textit{target} images on which we vary three different parameters:

\begin{enumerate}
    \item \textbf{Size:} The image size in pixels. We tested on square images with sizes $256\times256$, $512\times512$, $768\times768$, and $1024\times1024$ pixels.
    \item \textbf{Stars:} The total number of stars in the image. We tested fields with 300, 500, 1,000, and 10,000 stars.
    \item \textbf{Background Noise:} We vary the $\lambda$ parameter for the simulated Poisson noise of the CCD. We create images with $\lambda$ equal to 100, 500, 1,000, and 5,000 counts.
\end{enumerate}

With these parameters, we simulated 10 pairs of \textit{source} and \textit{target} images for each one of the 64 possible combinations, giving a total of 640 test cases.
The benchmark was then executed on a computer with the following specifications:


\begin{description}
    \item[CPU] -- 2.4GHz Intel Core i7-4700HQ (Quad-core, 6MB cache, up to 3.4GHz with Turbo Boost)
    \item[RAM] -- 16GB DDR3L (2x 8GB, 1,600MHz)
    \item[OS] -- Ubuntu Linux 18.04 64bits.
    \item[Software] -- \textit{Python} 3.7.3, \textit{NumPy} 1.16.4, \textit{SciPy} 1.3.0, \textit{Scikit-Image} 0.15.0 and \textit{Sep} 1.0.3
\end{description}

\begin{figure}
    \centering
    \includegraphics[width=.75\columnwidth]{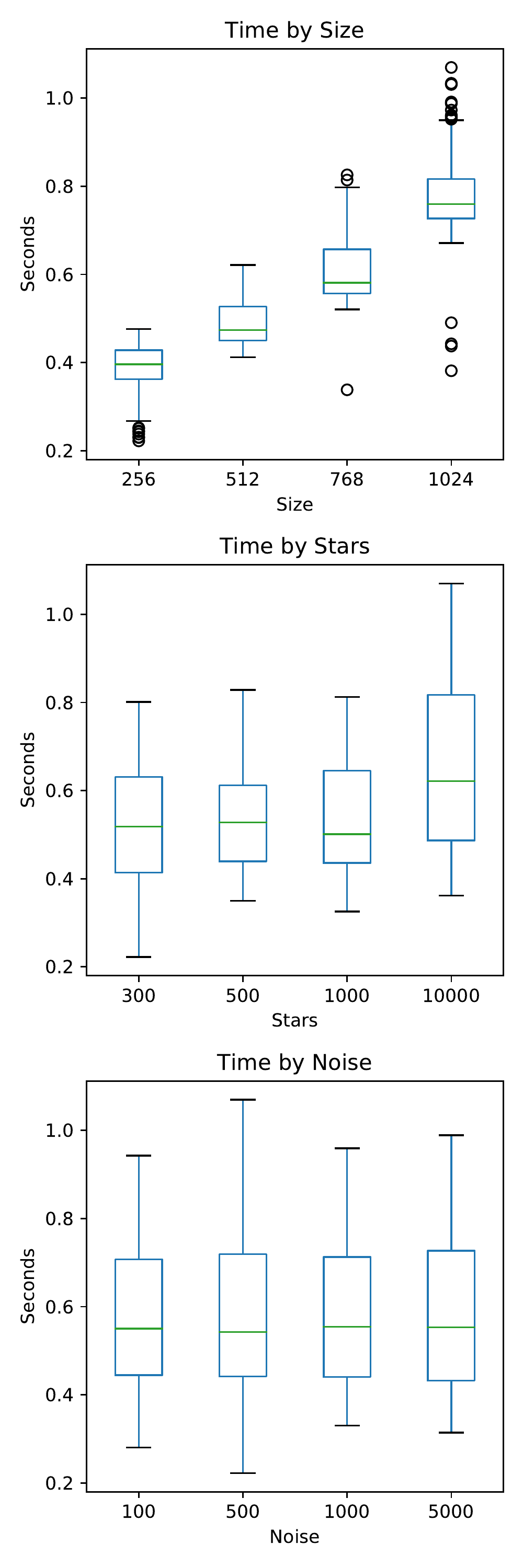}
    \caption{\label{fig:profile} Results of the benchmark on $640$ test cases varying the values of size, stars and background noise. In all cases, the vertical axis represents the execution time in seconds, while the horizontal axis represents the different parameter values.
    %
    %
    We can see that the algorithm increases execution time as the size of the images increases, while in all other cases the times remain relatively unchanged. This entire benchmark data-set can be found at: \url{https://github.com/toros-astro/astroalign/blob/master/benchmarks/time_benchmarks/20191023.csv}}
\end{figure}

The results show that the critical factor in determining the execution speed of a project is dependent on the size of the image to be registered (see Figure~\ref{fig:profile}).

\begin{figure}
    \centering
    \includegraphics[width=.95\columnwidth]{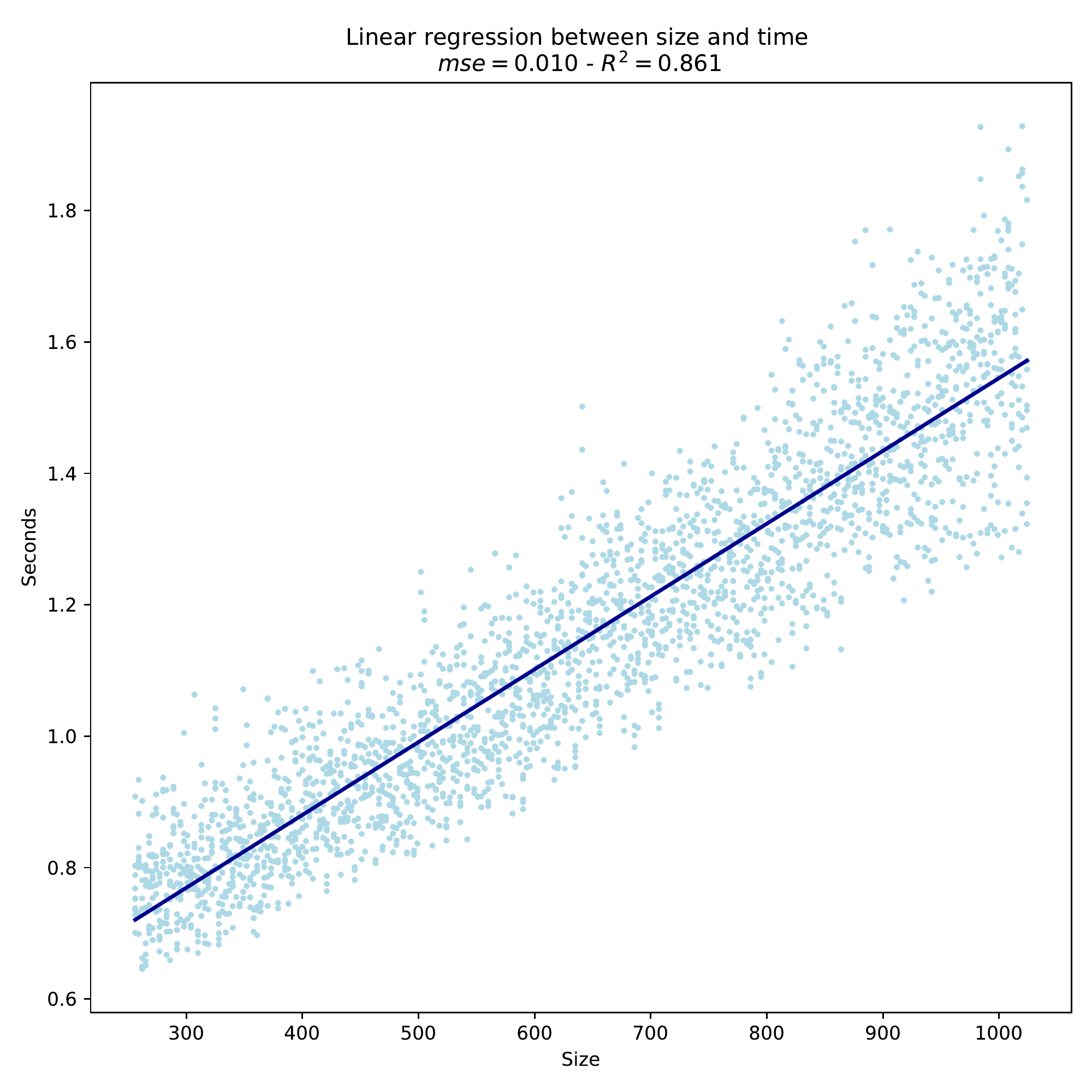}
    \caption{\label{fig:lrprofile}
    Measured registration times for $2,500$ pairs of random square images of sizes between $256 \times 256$ and $1024 \times 1024$ pixels. The images keep the amount of stars and background noise fixed at $10,000$ sources and $\lambda=1,000$ counts respectively.
    The horizontal axis represents the size of the image in pixels, and the vertical axis represents the registration time in seconds. Each dot is a measurements, and the blue line is the ordinary least squares regression line.
    This entire benchmark data-set can be found at:
    \url{https://github.com/toros-astro/astroalign/blob/master/benchmarks/time_regressions/20200220.csv}
    }
\end{figure}

To achieve more accurate results in the special case of time vs. size graph, we performed a second benchmark with $2,560$ pairs of random square images of sizes between $256 \times 256$ and $1024 \times 1024$. We kept the amount of stars and background noise fixed at $10,000$ sources and $\lambda=1,000$ counts respectively.
We can see in Figure~\ref{fig:lrprofile} that a linear regression fit adjusts well the data points. The regression has a mean squared error $MRE \sim 1\%$ and a coefficient of determination $R^2 \sim 0.861$.
The result of this second experiment seems to indicate that the algorithm has a linear growth in execution time as the size of the image increases.

The project running time is thus bounded and predictable, since it is usually the case that images in a processing pipeline are equal in size.

\subsection{Quality assurance}

Software quality assurance deals with the measurement of
qualitative and quantitative metrics. 
The most common techniques to extract this information are \textit{unit-testing} and \textit{code-coverage}. 

The objective of \textit{unit-testing} is to isolate and show that each part of the program is correct \citep{jazayeri_trends_2007}.
\textit{Code-coverage} is a measure of how much code is executed by the unit tests, expressed as a percentage \citep{miller1963systematic}.
Having an extensive code-coverage prevents restricting tests to only a subset of the entire project.
In the Astroalign project we provide a collection of nine unit-tests that complete $97\%$ of code-coverage using Python versions $2.7$, $3.6$ and $3.7$ \footnote{These metrics are calculated as of June 4th, 2019.}.

We are interested in the maintainability of the project and, for this reason, we endorse the PEP 8 -- Style Guide for Python Code \citep{van2001pep}, using the \textit{flake8} \footnote{\url{http://flake8.pycqa.org}} tool, which automatically checks for any deviation in style making code easier to understand \citep{latte2019clean}.
 
Finally, the entire source code is MIT-licensed \citep{open2006license}, and available in a public repository\footnote{\url{https://github.com/toros-astro/astroalign}}.
All versions committed to this code are automatically tested with a continuous-integration service \footnote{\url{https://travis-ci.org/toros-astro/astroalign}}, and documentation is automatically built from the repository and made public in the \textit{read-the-docs}  service\footnote{\url{https://astroalign.readthedocs.io/en/latest}} \citep{holscher_2016}.

With these techniques and tools we try to provide high quality for the Astroalign project and its development process.

\subsection{Integration with the Python scientific--stack}

Python is a very accessible language for casual programmers, including astronomers, and has become the tool of choice for almost every new astronomical project \citep{greenfield2011python}.
When the \textit{astropy} \citep{robitaille2013astropy} library was released, the entire astronomical community was able to build an entire astronomy-related ecosystem upon this same foundation \citep{price2018astropy}.
Therefore it is natural for Astroalign to also try to take advantage of this library.
The most relevant of these projects are available as astropy-affiliated projects\footnote{\url{https://www.astropy.org/affiliated/}}.

Astroalign is built on top of the Python scientific stack \textit{Numpy} \citep{van2011numpy} for modeling the image as an efficient multidimensional array; \textit{Scipy} \citep{jones2014scipy} for an implementation of k-d trees;  \textit{Scikit-Image} \citep{van2014scikit} for the basic image processing; and \textit{Source Extraction and Photometry (SEP)} \citep{barbary2016sep, bertin1996sextractor} for segmentation and analysis of astronomical images from a photometric point of view. 

Finally, Astroalign is available for installation on the \textit{Python-Package-Index} (PyPI)\footnote{\url{https://pypi.org/project/astroalign/}} and can be installed using the \textit{pip} program with the command \texttt{pip install astroalign}. 
This makes our project easy to integrate with any other tool in the Python ecosystem \citep{valiev2018ecosystem}.

Astroalign is registered with \textit{The Astrophysics Source Code Library (ASCL)} \citep{allen2014looking} and has been accepted as an Astropy affiliated project.


\section{Conclusion}

In this paper we presented Astroalign, a Python module to register stellar images, especially suited for the case where there is no WCS information available. Astroalign is well suited to use in astrophotography, coaddition, and image subtraction, and integrates well with other existing astronomy Python packages.

Image registration is done by identifying corresponding three-point asterisms (triangles) on both images and estimating the affine transformation between them. Asterism matching is a more robust and closer to the human way of matching images than generic image registration algorithms based on features.

Astroalign can reconcile images of very different field of view, point-spread function, seeing, and atmospheric conditions.

{\em The authors would like to acknowledge support from grants from the National Science Foundation of the United States of America, NSF HRD 1242090, and the Consejo Nacional de Investigaciones Cient\'{\i}ficas y T\'ecnicas (CONICET, Argentina).}

\section*{References}

\bibliographystyle{aa}
\bibliography{mybiblio}


\appendix

\section{Comparative Table for Registration Software Packages}
\label{appendix:a}

In the following lines we summarize some of the most recognized registration implementations.
See Section \ref{sec:comp} for a full discussion.

\small
\begin{itemize}
    \item {\normalsize Astroalign}
        \begin{description}
            \item[Language:] Python
            \item[Input type:] Images
            \item[Tessellation:] Nearest-neighbor triangulation
            \item[License:] \textsc{MIT}
            \item[Homepage:] \url{https://astroalign.readthedocs.io/}
            \item[Reference:]  \cite{martin_beroiz_2019_3406793, beroiz2019astroalign}
        \end{description}
        
    \item {\normalsize Astromety.net}
        \begin{description}
            \item[Language:] \textsc{ANSI-C}
            \item[Input type:] Images 
            \item[Tessellation:] Quads on Healpix cells
            \item[License:] \textsc{BSD 3-Clause}
            \item[Homepage:] \url{http://astrometry.net/}
            \item[Reference:] \cite{lang2010astrometry, lang2012astrometry}
        \end{description}
        
    \item {\normalsize \textsc{OPM\_A}}
        \begin{description}
            \item[Language:] \textsc{ANSI-C}
            \item[Input type:] Large lists of star positions with non-linear distortion 
            \item[Tessellation:] Triangulation with priority assigned to acute-angle triangles
            \item[License:] Custom
            \item[Homepage:] \url{http://stella.astron.s.u-tokyo.ac.jp/nmatsuna/Japanese/software/OPM.html} (Japanese)
            \item[Reference:] \citet{Tabur2007} 
        \end{description}
        
    \item {\normalsize \textsc{FITSH}} (\texttt{grmatch} and \texttt{grtrans})
        \begin{description}
            \item[Language:] \textsc{ANSI-C}
            \item[Input type:] Large lists of star positions
            \item[Tessellation:] Delaunay Triangulation \citep{delaunay1934sphere}
            \item[License:] \textsc{GPL-3}
            \item[Homepage:] \url{https://fitsh.net/}
            \item[Reference:] \cite{PalBakos2006, pal2012fitsh}
        \end{description}
\end{itemize}

\end{document}